# Maxwell Equations with Accounting of Tensor Properties of Time

Vlokh R. and Kvasnyuk O.

Institute of Physical Optics, 23 Dragomanov St., 79005 Lviv, Ukraine, e-mail: vlokh@ifo.lviv.ua

**Abstract**

The Maxwell equations with accounting for tensors properties of time have been considered. The effects that follow from such consideration are described. These are the appearance of vacuum polarization, anisotropy of electromagnetic wave velocity in vacuum, anisotropy of the vacuum dielectric permittivity, rotation of light polarization plane, as well as the existence of longitudinal components of electromagnetic wave and the rotational (non-potential) component of electric field caused by electric charges.

## Introduction

One of the commonly used notations of Maxwell equations for vacuum is as follows (see, e.g., [1]):

$$\begin{cases} \varepsilon_0 \frac{\partial E}{\partial t} = rotH \\ -\mu_0 \frac{\partial H}{\partial t} = rotE \\ divE = 0 \\ divH = 0 \end{cases} \qquad (1)$$

where $E$ and $H$ are respectively polar and axial vectors of electric and magnetic fields, $t$ is time represented by scalar quantity and $\varepsilon_0$ and $\mu_0$ stand respectively for electric permittivity and magnetic permeability of vacuum. The first two relations in Eq. (1) are known as the Faraday's and Ampere's laws, respectively, while the last two as the Coulomb's and Gauss's laws. In order to describe electromagnetic wave propagation in material media with certain electric and magnetic properties, the relations (1) are supplemented with the constitutive relations

$$\begin{cases} J + \varepsilon \dfrac{\partial E}{\partial t} = rotH \\ -\mu \dfrac{\partial H}{\partial t} = rotE \\ divD = \rho \\ divB = 0 \end{cases} \quad , \tag{2}$$

$$D_i = \varepsilon_{ij} E_j = \varepsilon_0 E_j + P_j = (\delta_{ij} + {}^e\chi_{ij})\varepsilon_0 E_j$$
$$B_i = \mu_{ij} H_j = \varepsilon_0 H_j + M_j = (\delta_{ij} + {}^m\chi_{ij})\mu_0 H_j$$
$$J_i = \sigma_{ij} E_j$$

where $\delta_{ij}$ is the Kronecker symbol, $P_j$ and $M_j$ represent respectively polar and axial vectors of polarization and magnetization, $D_i$, $B_i$ - vectors of electric and magnetic induction, $\varepsilon_{ij}$ (${}^e\chi_{ij}$) and $\mu_{ij}$ (${}^m\chi_{ij}$) polar second-rank tensors of electric and magnetic permittivities (susceptibilities), $\sigma_{ij}$ is a polar second-rank electric conductivity tensor and, finally, $J_i$ denotes a polar current density vector, $\rho$ - electric charge density.

The Maxwell equations have been repeatedly expanded and reformulated in order to search for possibility of their "symmetrization", i.e. to find the solutions describing "magnetic charges". For example, Dirac [2] has postulated existence of magnetic monopoles, which have never been detected experimentally and has proposed the following notation of the Maxwell equations:

$$\begin{cases} \varepsilon \dfrac{\partial E}{\partial t} + J_e = rotH \\ \mu \dfrac{\partial H}{\partial t} + J_m = -rotE \\ \varepsilon divE = \rho_e \\ \mu divH = \rho_m \end{cases} , \tag{3}$$

Another examples are the approach by Munera-Guzma [3] based on the assumption that the current and charge densities are merely electromagnetic entities or the approach by Harmuth and Meyl [4-8] assuming that there exist no monopoles, neither electric nor magnetic, while the electric charges are then only secondary effects of electric and magnetic fields. Inomata [9], Rauscher [10] and Honig [11] have used imaginary axis for including imaginary magnetic charge and current into the Maxwell equation. Summarizing different approaches, one can find that no satisfactory "symmetrization" of the Maxwell equation has been found up to now. Though the electric charges are measurable, there is no strong evidence for detection of magnetic monopoles, in spite of continuous experiments (see, e.g., [12-18]). Thus, one comes to conclusion that no magnetic potential fields could be postulated and non-symmetric form of the Maxwell equations is still correct. Nevertheless, the facts of existence of magnetic monopoles are still mentioned in some works (one of the recent ones is [19]). In the present paper we will rewrite the Maxwell equations, assuming tensor properties of the time, in order to check a possibility of their "symmetrization".

On the other side, the changes in the refractive index of vacuum induced by the gravitation field, which have been mentioned in [20-25], seem to be described in frame of a quantum theory with accounting for virtual electron-positron pairs. However, for description of these phenomena by electrodynamic relations it seems to be necessary to introduce some dielectric constant, which differs from the free space one $\varepsilon_0$ and, moreover, possesses tensor properties. Namely, in the present paper we will argue that, after accounting for lowering of space symmetry and appearing of tensor properties of the time it becomes possible to describe these problems by electrodynamics equations. It is necessary to notice that the tensor character of the time can be a property of Early Universe, which could possess a lower symmetry [16].

In the recent report [20] it has been shown that, in frame of the weak-field gravity and optical-mechanical analogy approach [21-23] (or the polarized vacuum approach [24]), the refractive index $n$ and the optical-frequency impermeability $B_{ij} = \left(1/n^2\right)_{ij}$ perturbed by the gravitation field of spherically symmetric mass may be written as

$$n = 1 + 2\sqrt{\beta_{ij}M}\,(g^{1/2}), \qquad (4)$$

$$B_{ij} = 1 - 4\sqrt{\beta_{ij}M}\,(g^{1/2}), \qquad (5)$$

where $\beta_{ij} = G/c_0^4$, $c_0$ is the light speed in vacuum and $G$ the gravitation constant. The square root of the gravitation field strength (the so-called free-fall acceleration) $g^{1/2}$ describes a scalar action, which cannot lead to symmetry lowering for a medium. On the other side, according to the results of our analysis [25], the initial birefringence of anisotropic media can be changed by the action of gravitation fields, even in the case of spherically symmetric mass. Furthermore, one can assume that the gravitation field of non-spherical mass would lead to lowering of the space symmetry, appearance of its anisotropy and so to an optical birefringence [26, 27]. Besides, the anisotropy could also appear due to considerable gradients of some scalar parameter [28] (in our case, the gravitation field of spherically symmetric mass). Those phenomena might be treated as gradient parametric optical effect [29].

The quantities $\beta$ and $G$ in Eqs. (4) and (5) are constitutive coefficients of a 3D flat space (or the corresponding optical medium) [20] and they should therefore obey von Neumann principle. Hence, lowering of initially spherical symmetry of the space by the gravitation field or the other fields can lead to appearance of tensor properties of the coefficient $G$. Moreover, if $\beta$ and $G$ are constitutive coefficients, then it follows, e.g., from the relation for the Hubble constant $H^2 = \frac{8}{3}\rho_c\pi G$ (with $\rho_c$ being the critical density of the Universe), that the above constant (because of the relation $H = \frac{1}{t}$, this is true of the time, too) plays a role of property of the flat space in the model of optical medium. The existing fields lead to lowering of the space symmetry, and these lowered symmetry groups allow a subsequent reduction of symmetry of properties of the space, i.e., the symmetry of the time (or that of the Hubble and the gravitation constants). Therefore, due to the Curie principle, the symmetry group of the flat space should depend on the field configuration and, according to the Neumann symmetry principle, it should be a subgroup of the symmetry group of the time. In other words, lowering of the space (or polarizable vacuum) symmetry under the action of gravitation field should lead to appearance of tensor properties of the gravitation coefficient *G*, the Hubble constant *H* and the time *t* [20], i.e. lowering of the time symmetry from a scalar to a second-rank tensor one. Notice that anisotropy (or deviation) of the gravitation constant from its Newtonian value is a subject of extensive theoretical and experimental investigations (see, e.g., [30-34]). The latter are additionally stimulated at present time by the observation of acceleration of the spacecrafts Pioneer 10 and 11 [35]. It is interesting to remind that the antisymmetric (AS) part of the Hubble constant has been recently used for describing cosmic velocities field in large scale [36]. Such a generalization of the Hubble constant seems to be followed by a similar generalization of the time.

In the present paper we will show how the Maxwell equations can be modified with taking tensor properties of the time into account and which conclusions follow from this approach.

## 2. Theoretical approach

Let us consider a second-rank polar tensor of the time as asymmetric tensor $t_{ij}$, which can be divided into the symmetric part,

$$T_{ij}^{s} = \frac{1}{2}(t_{ij} + t_{ji}), \quad (6)$$

and the AS one,

$$T_{ij}^{as} = \frac{1}{2}(t_{ij} - t_{ji}). \quad (7)$$

The symmetric part consists of a spherical ($\frac{1}{3}t_{mm}\delta_{jk}$) and a deviation ($t_{ij}^{D} = T_{ij}^{s} - \frac{1}{3}t_{mm}\delta_{ij}$) parts. It is the spherical part of the time tensor, in fact a scalar, characterizes processes in the isotropic space.

We will study the Maxwell equations for electromagnetic field, while considering the time as a second-rank tensor in the Euclidian space. Let us define the time tensor as a quantity that can be reduced to traditional definition of the time, i.e. a quantity that can be reduced to the scalar time in the non-degenerated isotropic space or the time as one of coordinate axes in the Minkowski space. As an example, let us consider two neighbouring volumes of the space. In one of which (namely, *B*) the anisotropy appears in the moment of time $t_0$, while the volume *A* remains isotropic. The observer situated in the volume *A* uses a scalar time $t_{mm}$ for describing of processes that take place in the neighbouring anisotropic volume *B*. Notice that, according to the Neumann symmetry principle, the anisotropy of the volume *B* should permit the appearance of tensor properties of the time. For describing the time as a second-rank tensor, the observer situated in the volume *A* should establish a functional dependence of each tensor component of the time on the scalar time which is a property of the volume $A$. Thus, we write the time tensor as

$$t_{ij} = \begin{vmatrix} t^A_{mm} + t^A_{mm}\Delta t_{11} & t^A_{mm}\Delta t_{12} & t^A_{mm}\Delta t_{13} \\ -t^A_{mm}\Delta t_{12} & t^A_{mm} + t^A_{mm}\Delta t_{22} & t^A_{mm}\Delta t_{23} \\ -t^A_{mm}\Delta t_{13} & -t^A_{mm}\Delta t_{23} & t^A_{mm} + t^A_{mm}\Delta t_{33} \end{vmatrix} = t^A_{mm} \begin{vmatrix} 1+\Delta t_{11} & \Delta t_{12} & \Delta t_{13} \\ -\Delta t_{12} & 1+\Delta t_{22} & \Delta t_{23} \\ -\Delta t_{13} & -\Delta t_{23} & 1+\Delta t_{33} \end{vmatrix} = t^A_{ij}(I + T^s_{ij} + T^{as}_{ij}), \qquad (8)$$

where the diagonal components include the unit tensor $I$, the symmetric tensor $\Delta t_{ij} = \Delta t_{ji} = T^s_{ij}$ and the AS tensor $\Delta t_{ij} = -\Delta t_{ji} = T^{as}_{ij}$, with $t^A_{mm}$ denoting the spherical part of the tensor (i.e., scalar). The equation for the electric component of electromagnetic wave may be rewritten with Eq. (8) as

$$\vec{E} = \vec{A}\exp(i\omega t^A_{mm}(I + T^s_{ij} + T^{as}_{ij}))\exp(ikr), \qquad (9)$$

where $\omega$ is the frequency, $k$ the wave vector and $r$ the radius vector. Then the partial derivative would be defined as follows:

$$\frac{\partial}{\partial[t^A_{mm}(I + T^s_{ij} + T^{as}_{ij})]} = \frac{1}{(I + T^s_{ij} + T^{as}_{ij})}\frac{\partial}{\partial t^A_{mm}}. \qquad (10)$$

### 3. Results and discussion
*a) Ampere's and Faraday's laws*

Replacing the time in the Maxwell equations according to $t \to t^A_{mm}(I + T^s_{ij} + T^{as}_{ij})$ and using Eqs. (9) and (10), one can obtain the two first Maxwell equations with accounting for tensor properties of the time:

$$\begin{cases} (rotE)_i + \dfrac{1}{(I+T^s_{ij}+T^{as}_{ij})}\dfrac{\partial B_j}{\partial t^A_{mm}} = 0 \\ (rotH)_i - \dfrac{1}{(I+T^s_{ij}+T^{as}_{ij})}\dfrac{\partial D_j}{\partial t^A_{mm}} = 0 \end{cases}. \qquad (11)$$

Let us study the behaviour of electromagnetic wave represented in the form of Eq. (9) for the case when only symmetric part of the time tensor $T^s_{ij}$ exists:

$$\vec{E} = \vec{e}\exp(i\omega t^A_{mm}(I + T^s_{ij}))\exp(ikr). \qquad (12)$$

Let us take into account that the exponential function of the matrix can be presented as [37]

$$\exp(i\omega t^A_{mm}(I + T^s_{ij})) = \sum_{n=0}^{\infty}\frac{(i\omega t^A_{mm}(I+T^s_{ij}))^n}{n!} = \begin{vmatrix} \exp(i\omega t^A_{mm}(1+T^s_{11})) & 0 & 0 \\ 0 & \exp(i\omega t^A_{mm}(1+T^s_{22})) & 0 \\ 0 & 0 & \exp(i\omega t^A_{mm}(1+T^s_{33})) \end{vmatrix}. \qquad (13)$$

Then Eq. (12) may be transformed to

$$\vec{E} = \vec{e} \begin{vmatrix} \exp(i\omega t_{mm}^A(1+T_{11}^s)) & 0 & 0 \\ 0 & \exp(i\omega t_{mm}^A(1+T_{22}^s)) & 0 \\ 0 & 0 & \exp(i\omega t_{mm}^A(1+T_{33}^s)) \end{vmatrix} \exp(ikr) =$$

$$= \begin{vmatrix} e_x \\ e_y \\ e_z \end{vmatrix} \begin{vmatrix} \exp(i\omega t_{mm}^A(1+T_{11}^s)) & 0 & 0 \\ 0 & \exp(i\omega t_{mm}^A(1+T_{22}^s)) & 0 \\ 0 & 0 & \exp(i\omega t_{mm}^A(1+T_{33}^s)) \end{vmatrix} \exp(ikr) = . \qquad (14)$$

$$= \begin{vmatrix} e_x \exp(i\omega t_{mm}^A(1+T_{11}^s))\exp(ikr) \\ e_y \exp(i\omega t_{mm}^A(1+T_{22}^s))\exp(ikr) \\ e_z \exp(i\omega t_{mm}^A(1+T_{33}^s))\exp(ikr) \end{vmatrix} = \begin{vmatrix} e_x \exp i(\omega t_{mm}^A(1+T_{11}^s)+kr) \\ e_y \exp i(\omega t_{mm}^A(1+T_{22}^s)+kr) \\ e_z \exp i(\omega t_{mm}^A(1+T_{33}^s)+kr) \end{vmatrix}$$

If the phase is constant, one can write the following system of equations:

$$\begin{cases} (\omega t_{mm}^A(1+T_{11}^s)+kr) = const \\ (\omega t_{mm}^A(1+T_{22}^s)+kr) = const \\ (\omega t_{mm}^A(1+T_{33}^s)+kr) = const \end{cases} . \qquad (15)$$

After derivation of both sides of these equations ($\frac{d}{dt_{mm}^A}$) we obtain

$$\begin{cases} (\omega(1+T_{11}^s)+kv_x) = 0 \\ (\omega(1+T_{22}^s)+kv_y) = 0 , \\ (\omega(1+T_{33}^s)+kv_z) = 0 \end{cases} \qquad (16)$$

and so the phase velocities $v_i$ of the electromagnetic wave are given by

$$\begin{cases} v_x = c_0(1-T_{11}^s) \\ v_y = c_0(1-T_{22}^s) \\ v_z = c_0(1-T_{33}^s) \end{cases} . \qquad (17)$$

From Eq. (17) it follows that the increment including $T_{ij}^s$ should be only negative, otherwise the phase velocity would exceed the light speed $c_0$ in the isotropic vacuum. It is worthwhile that the above conclusion is in agreement with the time delay appearing in the gravitation field that follows from the Lorentz transformations. Since the sign of all the components $T_{ij}^s$ in Eq. (17) is the same, this leads to important property of $T_{ij}^s$: $\mathrm{Sp} T_{ij}^s \neq 0$. Hence, the tensor properties of the time lead to anisotropy in the phase velocity of electromagnetic waves, similarly to anisotropic crystals with the refractive indices

$$\begin{cases} n_x = \dfrac{1}{(1-T_{11}^s)} \simeq 1+T_{11}^s \\ n_y = \dfrac{1}{(1-T_{11}^s)} \simeq 1+T_{11}^s, \\ n_z = \dfrac{1}{(1-T_{11}^s)} \simeq 1+T_{11}^s \end{cases} \qquad (18)$$

or the dielectric permittivity of free space:

$$\begin{cases} \varepsilon_{0x} = \dfrac{\varepsilon_0}{(1-T_{11}^s)^2} \simeq \dfrac{\varepsilon_0}{1-T_{11}^s} \\ \varepsilon_{0y} = \dfrac{\varepsilon_0}{(1-T_{11}^s)^2} \simeq \dfrac{\varepsilon_0}{1-T_{22}^s} . \\ \varepsilon_{0z} = \dfrac{\varepsilon_0}{(1-T_{33}^s)^2} \simeq \dfrac{\varepsilon_0}{1-T_{33}^s} \end{cases} \qquad (19)$$

Assume now that the time tensor includes a real AS part $T_{ij}^{as}$ (e.g., $T_{12} = -T_{21}$) and so we have

$$\begin{vmatrix} E_x \\ E_y \\ E_z \end{vmatrix} = \exp\begin{vmatrix} 0 & i\omega t_{mm}^A T_{ij}^{as} & 0 \\ -i\omega t_{mm}^A T_{ij}^{as} & 0 & 0 \\ 0 & 0 & 0 \end{vmatrix} \times \begin{vmatrix} e_x \\ e_y \\ e_z \end{vmatrix} = \begin{vmatrix} \cosh \omega t_{mm}^A T_{ij}^{as} & i\sinh \omega t_{mm}^A T_{ij}^{as} & 0 \\ -i\sinh \omega t_{mm}^A T_{ij}^{as} & \cosh \omega t_{mm}^A T_{ij}^{as} & 0 \\ 0 & 0 & 1 \end{vmatrix} \times \begin{vmatrix} e_x \\ e_y \\ e_z \end{vmatrix}, \qquad (20)$$

where the matrix functions may be found, e.g., in [37]. Then such the property of the time would lead to infinite increase in the electric field components, a physically impossible fact. If $T_{ij}^{as}$ is an inverse-proportional function of $t_{mm}^A$ and $\lim\limits_{t_{mm}^A \to \infty} \dfrac{T_{ij}^{as}}{1/t_{mm}^A} = 0$, the r. h. s. matrix in Eq. (20) would tend to the spherical matrix according to relation

$$\begin{vmatrix} E_x \\ E_y \\ E_z \end{vmatrix} = \lim_{t_{mm}^A \to \infty} \begin{vmatrix} \cosh \omega t_{mm}^A T_{ij}^{as} & i\sinh \omega t_{mm}^A T_{ij}^{as} & 0 \\ -i\sinh \omega t_{mm}^A T_{ij}^{as} & \cosh \omega t_{mm}^A T_{ij}^{as} & 0 \\ 0 & 0 & 1 \end{vmatrix} \times \begin{vmatrix} e_x \\ e_y \\ e_z \end{vmatrix} = \begin{vmatrix} 1 & 0 & 0 \\ 0 & 1 & 0 \\ 0 & 0 & 1 \end{vmatrix} \times \begin{vmatrix} e_x \\ e_y \\ e_z \end{vmatrix}, \qquad (21)$$

since one has

$$\lim_{t_{mm}^A \to \infty} \cosh \omega t_{mm}^A T_{ij}^{as} = 1 \\ \lim_{t_{mm}^A \to \infty} \sinh \omega t_{mm}^A T_{ij}^{as} = 0 . \qquad (22)$$

As seen from the above relations, spontaneous degeneration of the time tensor should vanish with increasing time scalar due to a reciprocal-power functional dependence, i.e. such a spontaneous AS-like lowering of symmetry of the space and time should disappear with time. For example, it can be a property of the early Universe. However, if the AS part of the time tensor is purely imaginary ($T_{ij}^{as} = \operatorname{Im} T_{ij}^{as}$), the electric field vector reads as

$$\begin{vmatrix} E_x \\ E_y \\ E_z \end{vmatrix} = \exp \begin{vmatrix} 0 & \omega t^A_{mm} T^{as}_{ij} & 0 \\ -\omega t^A_{mm} T^{as}_{ij} & 0 & 0 \\ 0 & 0 & 0 \end{vmatrix} \begin{vmatrix} e_x \\ e_y \\ e_z \end{vmatrix} = \begin{vmatrix} \cos(\omega t^A_{mm} T^{as}_{ij}) & \sin(\omega t^A_{mm} T^{as}_{ij}) & 0 \\ -\sin(\omega t^A_{mm} T^{as}_{ij}) & \cos(\omega t^A_{mm} T^{as}_{ij}) & 0 \\ 0 & 0 & 1 \end{vmatrix} \times \begin{vmatrix} e_x \\ e_y \\ e_z \end{vmatrix}. \quad (23)$$

Eq. (23) represents rotation of the vector $\vec{E}$ with respect to the vector $\vec{e}$ around Z axis by the angle

$$\varphi_z = \omega t^A_{mm} T^{as}_{ij}. \quad (24)$$

Actually, Eqs. (23) and (24) describe the gyration effect in the rotating space, whose symmetry permits the existence of AS part of the dielectric permittivity

$$\varepsilon_{0ij} = \frac{\varepsilon_0}{1 + T^s_{ij} + iT^{as}_{ij}}. \quad (25)$$

For instance, if we account for, e.g., the rotation around Z axis only, the $\varepsilon_{0ij}$ tensor in the matrix representation will take the form

$$\varepsilon_{0ij} = \begin{vmatrix} \dfrac{\varepsilon_0}{1 + T^s_{11}} & \dfrac{i\varepsilon_0 T_{12}(1 + T^s_{33})}{(1 + T^s_{11})(1 + T^s_{22})(1 + T^s_{33})} & 0 \\ -\dfrac{i\varepsilon_0 T_{12}(1 + T^s_{33})}{(1 + T^s_{11})(1 + T^s_{22})(1 + T^s_{33})} & \dfrac{\varepsilon_0}{1 + T^s_{22}} & 0 \\ 0 & 0 & \dfrac{\varepsilon_0(1 + T^s_{11})(1 + T^s_{22}) - T^2_{12}}{(1 + T^s_{11})(1 + T^s_{22})(1 + T^s_{33})} \end{vmatrix}. \quad (26)$$

Notice here that the Hermitian property of the dielectric permittivity tensor (i.e., purely imaginary character of the AS part of this tensor) is not postulated here though it follows logically from Eqs. (20)–(26). It is interesting to remind in this respect that an imaginary part of the metric tensor has been introduced in the works by Einstein (see, e.g., [38, 39]) for development of unified field theory. Such the metric tensor obeys the Hermitian principle ($g_{lk} = g^*_{kl}$) and its AS part describes a rotation of coordinate frame occurring during a parallel shift.

The existing of nonzero AS part of the time tensor leads to interesting property of divergence of electromagnetic field components. For example, if the electromagnetic wave propagates along Y axis and the amplitude of the electric component vector is represented as $\begin{vmatrix} e_x \\ 0 \\ e_z \end{vmatrix}$, the resulting wave may be written as

$$\begin{vmatrix} E_x \\ E_y \\ E_z \end{vmatrix} = \begin{vmatrix} \cos(\omega t^A_{mm} T^{as}_{ij}) & \sin(\omega t^A_{mm} T^{as}_{ij}) & 0 \\ -\sin(\omega t^A_{mm} T^{as}_{ij}) & \cos(\omega t^A_{mm} T^{as}_{ij}) & 0 \\ 0 & 0 & 1 \end{vmatrix} \times \begin{vmatrix} e_x \\ 0 \\ e_z \end{vmatrix} \exp i(\omega t^A_{mm} - k_y y) = \begin{vmatrix} e_x \cos \omega t^A_{mm} T^{as}_{ij} \\ -e_x \sin(\omega t^A_{mm} T^{as}_{ij}) \\ e_z \end{vmatrix} \exp i(\omega t^A_{mm} - k_y y). \quad (27)$$

The divergence of Eq. (27) is as follows:

$$div \begin{vmatrix} E_x \\ E_y \\ E_z \end{vmatrix} = iA_x k_y (\sin \omega t_{mm}^A T_{ij}^{as}) \exp i(\omega t_{mm}^A - k_y y). \quad (28)$$

One can see from Eqs. (27) and (28) that the incident wave propagating from isotropic space in the Y direction, with the amplitude components $e_x$ and $e_z$, acquires longitudinal components after crossing a boundary between the isotropic and anisotropic spaces.

As a consequence of our analysis and consideration of properties of the time tensor, we rewrite the first two Maxwell equations (Faraday's and Ampere's laws) as

$$rot\vec{E} + \frac{1}{1+T_{ij}^s + iT_{ij}^{as}} \frac{\partial \vec{B}}{\partial t_{mm}^A} = 0$$

$$rot\vec{H} - \frac{1}{1+T_{ij}^s + iT_{ij}^{as}} \frac{\partial \vec{D}}{\partial t_{mm}^A} = 0 \quad (29)$$

It is easy to check that the wave given by Eq. (12) is a solution of Eqs. (29).

*b) Coulomb's law*

In the present paper we will consider the Coulomb's law only in brief. Let us write out the electric field created by a charge $q$, which is introduced into the space with the tensor time properties, in the following form:

$$E_i = \frac{q}{4\pi r^3 \varepsilon_0} r_i = \frac{q}{4\pi r^3 \varepsilon_0} (I + T_{ij}^s + T_{ij}^{as}) r_j, \quad (30)$$

according to Eq. (21), or

$$E_i = \frac{q}{4\pi r^3 \varepsilon_0} r_i = \frac{q}{4\pi r^3 \varepsilon_0} (I + T_{ij}^s + iT_{ij}^{as}) r_j, \quad (31)$$

according to Eq. (23).

Due to accounting for symmetric part of the time tensor, the electric field strength caused by the electric charge should be anisotropic in the both cases. However, according to the duality condition, AS polar second-rank tensors may be rewritten as axial vectors ($\tau_l = -\frac{1}{2}\delta_{lij} T_{ij}^{as}$, where $\delta_{lij}$ denotes the Levi-Civita pseudo-tensor). Then the existence of AS part of the time tensor leads to the relations

$$E_i = -\frac{q}{2\pi r^3 \varepsilon_0} \delta_{lij} \left[ T_{ij}^{as} \times r_j \right] = -\frac{q}{2\pi r^3 \varepsilon_0} [\tau \times r], \quad (32)$$

according to Eq. (30), or

$$E_i = -i\frac{q}{2\pi r^3 \varepsilon_0} \delta_{lij} \left[ T_{ij}^{as} \times r_j \right] = -i\frac{q}{2\pi r^3 \varepsilon_0} [\tau \times r], \quad (33)$$

according to Eq. (31). It is obvious that Eq. (32) describes some kind of vortex, which is created by electric charges. The rotational component of the electric field does not represent a potential field, thus describing the action similar to that of the magnetic field. It is interesting to note that the magnetic monopoles are thought to arise in the early Universe in the grand unified theories. Most probably, the case of imaginary AS time tensor (see Eq. (33)) describes a rotation of polarization plane in such a kind of space, which permits nonzero AS time. This can be readily explained by the fact that the dielectric permittivity in Eq. (33) is AS and imaginary.

The divergence of the electric induction caused by the point charge in this anisotropic space can be presented as

$$divD_i = q\delta(r_i) + q\iiint_G -\delta(r_i)\frac{T^s_{11}x^2 + T^s_{22}y^2 + T^s_{33}z^2}{\varepsilon_o r^5} + \frac{\delta(r_i)(T^s_{11} + T^s_{22} + T^s_{33})}{\varepsilon_o r^3}dxdydz .\qquad(34)$$

Thus, if the charges are randomly distributed, the third Maxwell equation takes the form

$$divD_i(^1r_i) = \rho(^1r_1) + \iiint_G \rho(^2r_i)\left[-\frac{T^{s\ 2}_{11}x^2 + T^{s\ 2}_{22}y^2 + T^{s\ 2}_{33}z^2}{\varepsilon_o |^2r - {}^1r|^5} + \frac{(T^s_{11} + T^s_{22} + T^s_{33})}{\varepsilon_o |^2r - {}^1r|^3}\right]dx_2 dy_2 dz_2 ,\qquad(35)$$

where $^1r$ is the radius vector of the point where the divergence is calculated and $^2r$ the radius vector of the region with the coordinates $x_2 = -\infty \div +\infty$, $y_2 = -\infty \div +\infty$, $z_2 = -\infty \div +\infty$. As one can see, Eq. (35) is reduced to the common Maxwell equation ($divD_i(^1r_i) = \rho(^1r_1)$), if $T^s_{11} = T^s_{22} = T^s_{33} \neq 0$ or $T^s_{ij} = 0$.

The static polarization of vacuum with the tensorial time may be represented as (see Eq.26)

$$P_i = (\varepsilon_{0ij} - \varepsilon_0\delta_{ij})E_j .\qquad(36)$$

It is seen that the vacuum polarization would not vanish ($P_i = \frac{\varepsilon_0 T_{ij}}{1 - T_{ij}}E_j$) even if $T^s_{11} = T^s_{22} = T^s_{33} \neq 0$ and $T^{as}_{ij} = 0$. According to the present analysis, the coefficient $K$, which has been introduced via the relation $D = \varepsilon E = \varepsilon_0 KE$ in the work [24] (with $\varepsilon$ being the dielectric permittivity of the polarized vacuum), appears in its natural way: $(K)_{ij} = \frac{\varepsilon_{oij}}{\varepsilon_0}$.

A more detailed analysis of the mentioned effect, in particular on the basis of quantum electrodynamics, will be a subject of forthcoming papers.

**Conclusions**

We have rewritten the Maxwell equations with consideration of tensor properties of the time. It has been demonstrated that introduction of the time tensor enables describing the changes in the velocities of electromagnetic waves in the so-called polarized vacuum, as well as anisotropy of those velocities. The relations obtained by us could be simply reduced to the common equations in case if the deviation and the AS part of the time tensor tend to zero. It has been shown that the Hermitian principle for the dielectric permittivity tensor follows from stable solutions for the AS part of the time tensor. A rotation of polarization plane of electromagnetic waves should appear when the space symmetry is lowered in such a way that the AS part of the time tensor becomes nonzero. Moreover, the existence of AS part of the time tensor leads to possible appearance of longitudinal polarization of electromagnetic wave and, according to the modified Coulomb's law, the electrostatic force should become anisotropic. Real AS parts of the time tensor can lead to appearance of a vortex-like electric field. The rotational component of the electric field does not represent potential field, but rather describes some action, which is similar to that of the magnetic field. At the same time, purely imaginary AS time tensor most probably describes a rotation of polarization plane in such a space. We have also shown that the vacuum can become polarized in the space if anisotropy of the time does not exist. Even time delay caused by the gravitation field or some other fields could lead to polarization of the vacuum.

**Acknowledgement**

The authors acknowledge financial support of this study from the Ministry of Education and Science of Ukraine (the Project N0106U000615).

**References**

1. Landau L.D. and Lifshitz E.M. Theoretical physics. Field theory. Moscow: "Nauka" (1973).

2. Dirac PAM 1931. Quantised Singularities in the Electromagnetic Field, Proc. London Roy. Soc. **A133**: 60-72.

3. Munera Héctor A and Octavio Guzma, 1997. A Symmetric Formulation of Maxwell Equations. Mod. Phys. Lett. **A 12**: 2089-2101.

4. Meyl K, 1990. Potentialwirbel, Indel Verlag, Villingen-Schwenningen Band **1** ISBN 3-9802542-1-6.

5. Meyl K, 1992. Potentialwirbel, Indel Verlag, Villingen-Schwenningen Band **2** ISBN 3-9802542-2-4.

6. Harmuth Henning F, 1986. Corrections of Maxwell equations for signals I. IEEE Transactions of Electromagnetic Compatibility EMC-**28**: 250-258.

7. Harmuth Henning F, 1986. Corrections of Maxwell equations for signals II. IEEE Transactions of Electromagnetic Compatibility EMC-**28**: 259-266

8. Harmuth Henning F, 1988. Reply to T.W. Barrett's Comments on the Harmuth ansatz: Use of a magnetic current density in the calculation of the propagation velocity of signals by amended Maxwell theory, IEEE Transactions of Electromagnetic Compatibility EMC-**30**: 420-421.

9. Inomata Shiuji, Paradigm of New Science – Principals for the 21st Centur**y,** Gijutsu Shuppan Pub. Co. Ltd. Tokyo (1987).

10. Rauscher Elizabeth A, Electromagnetic Phenomena in Complex Geometries and Nonlinear Phenomena, Non-Hertzian Waves and Magnetic Monopoles, Tesla Book Company, Chula Vista CA-91912. (1983).

11. Honig William M, Quaternionic Electromagnetic Wave Equation and a Dual Charge-Filled Space. Lettere al Nuovo Cimento, Ser. 2 19 /4 (28 Maggio 1977) 137-140.

12. De Rujula A, 1995. Effects of virtual monopoles. Nucl. Phys. **B435**: 257-276.

13. G. 't Hooft, 1974. Magnetic monopoles in unified gauge theories. Nucl. Phys. **B79**: 276-284.

14. Polyakov AM, 1974. Particle Spectrum in Quantum Field Theory. JETP Lett. **20**: 194-195.

15. Craigie NS, Giacomelli G, Nahm W and Shafi Q, Theory and Detection of Magnetic Monopoles in Gauge Theories, World Scientific: Singapore (1986).

16. Lazarides G, Panagiotakopoulos C and Shafi Q, 1987. Magnetic monopoles from superstring models. Phys. Rev. Lett. **58**: 1707-1710.

17. Bhattacharjee P and Sigl G, 2000. Origin and Propagation of Extremely High Energy Cosmic Rays. Phys. Rept. **327**: 109-247.

18. Bertani M, Giacomelli G, Mondardini MR, Pal B, Patrizii L, Predieri F, Serralugaresi P, Sini G, Spurio M, Togo V and Zucchelli S, 1990. Search for magnetic monopoles at the Tevatron collider. Europhys. Lett. **12**: 613-616.

19. Fang Z, Nagaosa N, Kei S Takahashi, Asamitsu A, Mathieu R, Ogasawara T, Yamada H, Kawasaki M, Tokura Y and Terakura K, 2003. The Anomalous Hall Effect and Magnetic Monopoles in Momentum Space. Science **302**: 92–95.

20. Vlokh R, 2004. Change of optical properties of space under gravitation field. Ukr. J. Phys. Opt. **5**: 27-31.

21. Nandi KK and Islam A 1995. On the optical-mechanical analogy in general relativity. Amer. J. Phys. **63**: 251-256.

22. Evans J, Nandi KK and Islam A, 1996. The optical-mechanical analogy in general relativity: Exact Newtonian forms for the equations of motion of particles and photons. Gen. Rel. Grav. **28**: 413-438.

23. Fernando de Felice. 1971. On the gravitational field acting as an optical medium. Gen. Rel. Grav. **2**: 347.

24. Puthoff HE, 2002. Polarizable-Vacuum (PV) Approach to General Relativity. Found. Phys. **32**: 927-943.

25. Vlokh R and Kostyrko M, 2005. Estimation of the Birefringence Change in Crystals Induced by Gravitation Field. Ukr. J. Phys. Opt. **6**: 125-127.

26. Boonserm P, Cattoen C, Faber T, Visser M and Weinfurtner S, 2005. Effective Refractive Index Tensor for Weak-Field Gravity Class. Quant. Grav. **22:** 1905.

27. Vlokh R and Kostyrko M, 2006. Comment on "Effective Refractive Index Tensor for Weak-Field Gravity" by P. Boonserm, C. Cattoen, T. Faber, M. Visser and S. Weinfurtner. Ukr. J. Phys. Opt. **7:** 147.

28. Savchenko AYu and Zel'dovich B Ya, 1994. Birefringence by a smoothly inhomogeneous locally isotropic medium: Three-dimensional case. Phys. Rev. E. **50:** 2287-2292.

29. Vlokh R, 1991. Nonlinear medium polarization with account of gradient invariants. Phys. Stat. Sol. (b) **168:** K47-K50.

30. Nordtvedt K, 2003. *d*G/*dt* measurement and the timing of lunar laser ranging observations. Class. Quant. Grav. **20**: L147–L154.

31. Mansouri R, Nasseri F and Khorrami M, 1999. Effective time variation of universe with variable space dimension A 259 194: gr-qc/9905052.

32. Bronnikov KA, Kononogov SA and Melnikov VN, 2006. Brane world corrections to Newton's law. Gen. Relativ. Gravit. **38:** 1215–1232.

33. Will CM, 1971. Relativistic gravity in the Solar system. II. Anisotropy in the Newtonian gravitation constant. Astrophys. J. **169**: 141-155.

34. Quinn TJ, Speake CC, Richman SJ, Davis RS and Picard A, 2001. A New Determination of *G* Using Two Methods. Phys. Rev. Lett. **87:** 111101-111105.

35. Anderson JD, Laing PA, Lau EL, Liu AS, Nieto MM and Turyshev SG, 1998. Indication, from Pioneer 10/11, Galileo, and Ulysses Data, of an Apparent Anomalous, Weak, Long-Range Acceleration. Phys. Rev. Lett. **81:** 2858-2861.

36. Fliche HH, Souriau JM and Triay R, 2006. Anisotropic Hubble expansion of large scale structures. Gen. Relativ. Gravit. **38(3)**: 463–474.

37. Gantmacher F.R. Theory of matrices. Moscow: Nauka (1988).

38. Einstein A, 1945. Generalisation of the relativistic theory of gravitation. Ann. Math. 46: 578-584.

39. Einstein A and Kaufmann B, 1955. A new form of the general relativistic field equations Ann. Math. **62**: 128-138.